\def\be{\begin{equation}}
\def\ee{\end{equation}}
\def\ba{\begin{eqnarray}}
\def\ea{\end{eqnarray}}
\documentclass[preprint,12pt]{elsarticle}

\usepackage{graphics}
\usepackage{epsfig}
\usepackage{graphicx}
\usepackage{amsmath}
\usepackage{amscd}

\usepackage{amssymb}

\begin{document}

\begin{frontmatter}



\title{Dynamical fluctuations in classical adiabatic
processes: General description and their implications}


\author[label1,label2]{Qi Zhang}
\author[label3,label4]{Jiangbin Gong\corref{cor1}}
\ead{phygj@nus.edu.sg}

\fntext[l]{Department of Physics, 2 Science Drive 3, Singapore, 117542, phone 65-6516-1154, Fax 65-6777-6126}
\author[label2,label5]{C.H. Oh}

\address[label1]{Center for Clean Energy and Quantum Structures and School of Physics and Engineering, Zhengzhou
University, Zhengzhou, 450052, China}
\address[label2]{Centre for Quantum Technologies and Department of
Physics, National University of Singapore, 117543, Singapore}
\address[label3]{Department of Physics and Centre for Computational
Science and Engineering, National University of Singapore, 117542,
Singapore}
\address[label4]{NUS Graduate School for Integrative Sciences
and Engineering, Singapore 117597, Republic of Singapore}
\cortext[cor1]{*}
\address[label5]{Institute of Advanced Studies, Nanyang Technological
University, Singapore 639798, Republic of Singapore}

\begin{abstract}
Dynamical fluctuations in classical adiabatic processes are not
considered by the conventional classical adiabatic theorem.  In this
work a general result is derived to describe the intrinsic dynamical
fluctuations in classical adiabatic processes. Interesting
implications of our general result are discussed via two subtopics,
namely, an intriguing adiabatic geometric phase in a dynamical model
with an adiabatically moving fixed-point solution, and the possible
``pollution" to Hannay's angle or to other adiabatic phase objects for
adiabatic processes involving non-fixed-point solutions.
\end{abstract}


\end{frontmatter}






\section{Introduction}

The adiabatic theorem is important in both classical and quantum
mechanics \cite{adiabatic}. It predicts a system's dynamical
behavior subject to slowly varying system parameters. Although a
general and mathematically rigorous proof of the adiabatic theorem
is not obvious in both classical mechanics and quantum mechanics,
the adiabatic theorem has been widely used. Indeed, it is always
highly useful so long as there exist two drastically different time
scales.   The adiabatic theorem has also led to the discoveries of
Berry phase \cite{Berryphase} and the classical counterpart, i.e.,
Hannay's angle \cite{Hannay}.

We focus on the classical adiabatic theorem (CAT), but as shown
below, some of our results can be applied to quantum systems as
well. Our interest here is not in a rigorous proof of the CAT, but
in dynamical fluctuations around what is predicted by CAT.  As
discussed below, the possible consequences of the fluctuations
neglected by the conventional CAT can be far reaching.  The
motivation of considering the fluctuations is based on a simple
observation. That is, CAT, whose proof is based on an average over
fast-varying variables, only reflects a mean dynamical behavior.  As
such fluctuations on top of a mean dynamical behavior should exist
in classical adiabatic processes.  Though fluctuations should be
intuitively smaller in a slower adiabatic process, their effects are
accumulated over a longer time scale and hence might not vanish even
in the adiabatic limit. For instance, in a few early
studies~\cite{Golin1, Golin2, Berry1996Non}, including the study of
``Hannay's angle of the world"~\cite{Berry1996Non,adam}, the actual
total change in canonical variables may depend on the smoothness of
the evolving adiabatic parameters. This abnormal behavior was shown
to be connected with subtle fluctuations in the action variables
from their average behavior predicted by CAT. Clearly then,  a
general description of the dynamical fluctuations in adiabatically
evolving and classically integrable systems should be of importance.

We shall present in this work a general result that describes the
dynamical fluctuations inherent to classical adiabatic processes.
Roughly speaking, it establishes an interesting connection between
the actual rate of change of slowly varying system parameters and
the actual classical orbits deformed from that predicted by CAT. To
illustrate the usefulness of our general result, we design a simple
dynamical model with an adiabatically moving fixed-point solution,
from which an intriguing classical geometric phase can emerge.
We then exploit our general result to discuss the ``pollution" to
Hannay's angle in classical adiabatic processes. A mean-field model
that describes a two-mode Bose-Einstein condensate (BEC) is also
proposed to study fluctuation-induced ``pollution" to adiabatic
quantum evolution.

To tackle with dynamical fluctuations, one may quickly think of an
equation describing the time dependence of the fluctuations around
ideal adiabatic orbits.  But this approach may not be fruitful
because in principle, the time dependence of any canonical variables
is already fully captured by classical canonical equations of
motion.  Instead, we are concerned with how fluctuations distort
trajectories as compared with that predicted by CAT. In this sense,
our approach is somewhat in a similar spirit as an early
``multiple-time-scale-expansion" approach to corrections to
classical adiabatic invariants in chaotic systems \cite{jarzynski2}.
However, we focus on fluctuations associated with individual orbits
in integrable systems, rather than fluctuations associated with an
ensemble of chaotic trajectories in an energy shell.

This paper is organized as follows.  In Sec.~II we derive a
differential equation describing the dynamical fluctuations in
classical adiabatic processes. Some related details are also
provided in Appendix. As an application, in Sec.~III we study the
case of an adiabatically moving fixed-point solution and show how an
intriguing geometric angle may emerge in a simple toy model. Based
on our general result, Sec.~IV discusses why the ``pollution" to
Hannay's angle may exist and then proposes a physical system to
study analogous fluctuation-induced pollution.
We finally give a brief summary in Sec.~V.

\section{General Description of Dynamical Fluctuations in Classical Adiabatic Processes}

Consider a classical integrable system with $N$ degrees of freedom.
Its Hamiltonian is given by $H(\mathbf{p},\mathbf{q}, \mathbf{R})$,
where canonical variables $\mathbf{p}=(p_1,p_2,\ldots,p_N)$ and
$\mathbf{q}=(q_1,q_2,\ldots,q_N)$ represent canonical momenta and
coordinates, and $\mathbf{R}$ represents a collection of system
parameters. Let $F(\mathbf{I},\mathbf{q},\mathbf{R})$ be the
generating function that induces the $\mathbf{R}$-dependent
canonical transformation from $(\mathbf{p},\mathbf{q})$ to the
action-angle variables $(\mathbf{I},\mathbf{\Theta})$, where
$I_i=\frac{1}{2\pi}\oint p_i dq_i$ and
$\mathbf{\Theta}=(\theta_1,\theta_2,\ldots,\theta_N)$.

To clearly present our derivation of a differential equation that
describes dynamical fluctuations in classical adiabatic processes,
this section is divided into four subsections representing the four
steps in our derivation. First, after expressing classical equations
of motion in the action-angle variables
$(\mathbf{I},\mathbf{\Theta})$, we define dynamical fluctuations on
top of the idealized solution given by CAT. Second, the time
dependence of the canonical variables $(\mathbf{p},\mathbf{q})$ is
expressed in terms of the dynamical fluctuations we define. Third,
directly using the canonical equations of motion and the canonical
transformation between the action-angle variables and the canonical
variables, we reexpress the time dependence of the canonical
variables in terms of the dynamical fluctuations as well as the
action-angle variables along idealized classical orbits. Finally, by
comparing results in the second and third steps a differential
equation describing the dynamical fluctuations around idealized
adiabatic orbits is obtained.

\subsection{Dynamical fluctuations}

In the $(\mathbf{I},\mathbf{\Theta})$ representation an integrable
Hamiltonian becomes $\mathcal{H}(\mathbf{I},\mathbf{R})$, which is
independent of the angle variables $\mathbf{\Theta}$.  For
time-varying $\mathbf{R}=\mathbf{R}(t)$, the equations of motion for
$(\mathbf{I},\mathbf{\Theta})$ are given by~\cite{Berry1985JPA}
\begin{eqnarray} \label{new-dyna1}
\frac{d I_i}{dt}& = &-\frac{\partial \mathbf{W}}{\partial \theta_i}\cdot \frac{d\mathbf{R}}{dt}, \\
\label{new-dyna2} \frac{d\theta_i}{dt}& =
&\omega_i(\mathbf{I};\mathbf{R})+ \frac{\partial
\mathbf{W}}{\partial I_i}\cdot\frac{d\mathbf{R}}{dt} ,
\end{eqnarray}
where $\omega_i(\mathbf{I},\mathbf{R})=\partial\mathcal{H}/\partial
I_i$ is the angular frequency, and $\mathbf{W}$ is defined by
\begin{equation}
\mathbf{W} \equiv \mathbf{\nabla}_{\mathbf{R}}
F[\mathbf{I},\mathbf{q}(\mathbf{I},\mathbf{\Theta},\mathbf{R}),\mathbf{R}]-\mathbf{p}\cdot\mathbf{\nabla}_{\mathbf{R}}
\mathbf{q}(\mathbf{I},\mathbf{\Theta},\mathbf{R}).
 \end{equation}
Note that $\mathbf{\nabla}_{\mathbf{R}}$ refers to the gradient in
the parameter space under fixed $(\mathbf{I},\mathbf{\Theta})$. If
\begin{eqnarray}
\epsilon\equiv \left|\frac{d\mathbf{R}}{dt}\right|
\end{eqnarray}
 is much smaller than
$\omega_i|\mathbf{R}|$, one can take the average of
Eq.~(\ref{new-dyna1}) over the rapidly oscillating angle variables,
yielding $\frac{dI_i}{dt} \approx 0$ ($\mathbf{W}$ is a periodic
function of $\mathbf{\Theta}$). CAT hence identifies the action
variables as adiabatic invariants, i.e., in adiabatic processes
their values are fixed at
$\overline{\mathbf{I}}\equiv(\overline{I}_1,\overline{I}_2,\ldots,
\overline{I}_N)$.  For clarity, angle variables associated with this
idealized solution are defined as $\overline{\mathbf{\Theta}}\equiv
(\overline{\theta}_1,\overline{\theta}_2,\ldots,
\overline{\theta}_N)$. We also use $\overline{\mathbf{p}}\equiv
\mathbf{p}(\overline{\mathbf{I}},\overline{\mathbf{\Theta}},\mathbf{R})$
and $\overline{\mathbf{q}}\equiv
\mathbf{q}(\overline{\mathbf{I}},\overline{\mathbf{\Theta}},\mathbf{R})$
to describe the idealized solution in terms of the (old) set of
canonical variables. With the action variables fixed at
$\overline{\mathbf{I}}$, one may then solve Eq.~(\ref{new-dyna2})
for a cyclic process from $t=0$ to $t=T$ in a straightforward
manner.  One may further take the average of the idealized solution
over all possible initial angle values to obtain Hannay's angle,
which is the total mean angle change minus a dynamical angle.

The above discussion does not represent a complete description of
classical adiabatic processes.  Clearly, Eq. (\ref{new-dyna1}) tells
us that $\frac{dI_i}{dt}$ is not mathematically zero:
 it may possess fluctuations of the order $O(\epsilon)$ (i.e., to the first order of $\epsilon$).
 As such, in performing an averaging procedure as is done in CAT
 one neglects the dynamical correlation between $\mathbf{\Theta}$ and $\mathbf{I}$.
It is hence necessary to reconsider Eq.~(\ref{new-dyna1}) in order
to consider any possible real-orbit fluctuations on top of CAT. On a
real orbit we assume we have
  $I_i=\overline{I}_i+\delta{I}_i$, where we have used $\delta$ to represent fluctuations from
  the behavior predicted by CAT.
 Equivalent to that, one can describe the same fluctuations
from the idealized orbit in terms of $\delta q_j$ and $\delta p_j$.

There are now both idealized adiabatic orbits without considering
fluctuations and true orbits with fluctuations: the geometry of an
idealized orbit can be characterized by
$\mathbf{I}=\overline{\mathbf{I}}$ and
$\overline{\mathbf{\Theta}}\in[0,2\pi)$; and that of a true orbit
with fluctuations is slightly deformed to
\begin{eqnarray}
\mathbf{I}&=&\overline{\mathbf{I}}+\delta \mathbf{I}, \\
\mathbf{\Theta}&=&\overline{\mathbf{\Theta}}+\delta\mathbf{\Theta},
\end{eqnarray}
where $\delta \mathbf{I}$ and $\delta\mathbf{\Theta}$ are assumed to
be at most of the order $O(\epsilon)$. By our definitions above, we
have
\begin{equation} \label{deltadefine}
\left(\begin{array}{c}\delta \mathbf{I} \\
\delta \mathbf{\Theta}\end{array}\right)
 =\left(\begin{array}{c}\frac{\partial \overline{\mathbf{I}}}{\partial
\overline{p}_j}\delta p_j+\frac{\partial \overline{\mathbf{I}}}{\partial \overline{q}_j} \delta q_j  \\
\frac{\partial \overline{\mathbf{\Theta}}}{\partial
\overline{p}_j}\delta p_j+\frac{\partial
\overline{\mathbf{\Theta}}}{\partial \overline{q}_j}\delta
q_j\end{array}\right)\equiv \left(\begin{array}{c}K \\ M
\end{array}\right) \left( \begin{array} {c} \delta \mathbf{p} \\
\delta\mathbf{q}\end{array}\right).
\end{equation}
Here and in the following the summation convention for repeated
indices is adopted.  Equation~(\ref{deltadefine}) also defines two
$N\times 2N$ matrices $K$ and $M$, corresponding to the upper and
lower halves of a Jacobi matrix. Note that throughout we use
$\frac{\partial\bar{f}}{\partial\bar{x}}$ to indicate
$\frac{\partial f}{\partial x}$ evaluated at $x=\bar{x}$.

As will be seen below, it suffices to consider fluctuations of the
first order of $\epsilon$ because higher-order effects cannot be
accumulated with time.  We stress that the fluctuations are
intrinsic: they are nonzero so long as $\epsilon$ is not identically
zero. In other words, fluctuations considered here exist in any
classical adiabatic process and should not be thought of an effect
arising from a too-large $\epsilon$. It should be also noted that in
principle, all the dynamical information is contained in
Eqs.~(\ref{new-dyna1}) and (\ref{new-dyna2}).  However, we are
interested in developing a framework to describe how fluctuations
might behave along an idealized classical orbit.

\subsection{Canonical equations of motion in terms of fluctuations}

In terms of the fluctuations $\delta q_j$ and $\delta p_j$, we next
expand $H(\mathbf{p},\mathbf{q},\mathbf{R})$ around
$\overline{H}\equiv
H(\overline{\mathbf{p}},\overline{\mathbf{q}},\mathbf{R})$ to the
order $O(\epsilon)$, yielding the following canonical equations of
motion for $(\mathbf{q},\mathbf{p})$:
\begin{eqnarray} \label{deltapq}  \nonumber
\frac{d p_i}{dt}&=&-\frac{\partial \overline{H}}{\partial
\overline{q}_i}-\frac{\partial^2 \overline{H}}{\partial
\overline{q}_i \partial \overline{p}_j}\delta
p_j-\frac{\partial^2\overline{H}}{\partial \overline{q}_i \partial
\overline{q}_j} \delta q_j \\
&=& \frac{\partial \overline{p}_i}{\partial
\overline{\theta}_j}\omega_{j}(\mathbf{I},\mathbf{R})-\frac{\partial^2
\overline{H}}{\partial \overline{q}_i \partial \overline{p}_j}\delta
p_j-\frac{\partial^2\overline{H}}{\partial \overline{q}_i \partial
\overline{q}_j} \delta
q_j; \nonumber \\
\frac{d q_i}{dt}&=&\frac{\partial \overline{H}}{\partial
\overline{p}_i}+\frac{\partial^2\overline{H}}{\partial
\overline{p}_i \partial \overline{p}_j} \delta
p_j+\frac{\partial^2\overline{H}}{\partial \overline{p}_i \partial
\overline{q}_j}\delta q_j \nonumber \\
&=& \frac{\partial \overline{q}_i}{\partial \overline{\theta}_j}
\omega_{j}(\mathbf{I},\mathbf{R})+
\frac{\partial^2\overline{H}}{\partial \overline{p}_i \partial
\overline{p}_j} \delta p_j+\frac{\partial^2\overline{H}}{\partial
\overline{p}_i \partial \overline{q}_j}\delta q_j, \label{dqdp1}
\end{eqnarray}
where we have used the following two canonical relations
\begin{eqnarray}
 \frac{\partial \overline{I}_{j}}{\partial \overline{q}_i}&=&-\frac{\partial \overline{p}_i}{\partial \overline{\theta_j}};
 \nonumber \\
 \frac{\partial \overline{I}_{j}}{\partial \overline{p}_i}&=&\frac{\partial \overline{q}_i}{\partial \overline{\theta}_j}.
 \end{eqnarray}
Through Eq.~(\ref{dqdp1}) it is seen that the time dependence of the
canonical variables $(\mathbf{p},\mathbf{q})$ is connected to the
dynamical fluctuations $\delta q_j$ and $\delta p_j$, to the first
order of $\epsilon$.

\subsection{Time-dependence of canonical variables from action-angle variables}

The time evolution of the canonical variables
$(\mathbf{p},\mathbf{q})$ may be also directly obtained from the
canonical transformation from the action-angle variables to
$(\mathbf{p},\mathbf{q})$ and from the equations of motion given by
Eqs.~(\ref{new-dyna1}) and (\ref{new-dyna2}). In particular, using
\begin{eqnarray}
 \frac{d p_i}{dt}=\frac{\partial
p_i}{\partial \mathbf{R}}\frac{d\mathbf{R}}{dt}+ \frac{\partial
p_i}{\partial I_j}\frac{d I_j}{dt}+\frac{\partial p_i}{\partial
\theta_j}\frac{d\theta_j}{dt}
\end{eqnarray}
 and the analogous expression for
$\frac{d q_i}{dt}$, rewriting the derivatives in
Eqs.~(\ref{new-dyna1}) and (\ref{new-dyna2}) at
$(\mathbf{I},\mathbf{\Theta})$ in terms of those at
$(\overline{\mathbf{I}},\overline{\mathbf{\Theta}})$,  and
neglecting all terms that are at least $O(\epsilon^2)$, one arrives
at (see Appendix for details)
\begin{eqnarray}  \nonumber
\frac{d p_i}{dt}&=&\frac{\partial \overline{p}_i}{\partial
\mathbf{R}}\frac{d\mathbf{R}}{dt}-\frac{\partial
\overline{p}_i}{\partial \overline{I}_j}\frac{\partial \mathbf{W}
}{\partial \overline{\theta}_j}\cdot\frac{d\mathbf{R}}{dt}+
\frac{\partial \delta p_i}{\partial
\overline{\theta}_j}\omega_j(\bar{\mathbf{I}}, \mathbf{R}) \\
\nonumber & &+\ \frac{\partial \overline{p}_i}{\partial
\overline{\theta}_j}\left[\frac{\partial \mathbf{W}}{\partial
\overline{I}_j}\cdot
\frac{d\mathbf{R}}{dt}+\omega_j(\bar{\mathbf{I}},
\mathbf{R})+\frac{\partial \omega_j}{\partial \overline{I}_k}\delta
I_k\right]\\ \nonumber \frac{dq_i}{dt}&=& \frac{\partial
\overline{q}_i}{\partial \mathbf{R}}\frac{d\mathbf{R}}{dt}
-\frac{\partial \overline{q}_i}{\partial
\overline{I}_j}\frac{\partial \mathbf{W}}{\partial\overline{\theta}_j}\cdot\frac{d\mathbf{R}}{dt} + \frac{\partial \delta q_i}{\partial \overline{\theta}_j}\omega_j(\bar{\mathbf{I}}, \mathbf{R}) \\
&  &+\  \frac{\partial \overline{q}_i}{\partial
\overline{\theta}_j}\left[\frac{\partial \mathbf{W}}{\partial
\overline{I}_j}
\cdot\frac{d\mathbf{R}}{dt}+\omega_j(\bar{\mathbf{I}},\mathbf{R})+\frac{\partial
\omega_j}{\partial \overline{I}_k}\delta I_k\right]. \nonumber \\
\label{dqdp2}
\end{eqnarray}
Interestingly, due to the direct connection between
$(\mathbf{p},\mathbf{q})$ and $(\mathbf{I},\mathbf{\Theta})$, the
full time dependence of $(\mathbf{p},\mathbf{q})$ is connected with
dynamical fluctuations in a highly nontrivial manner. In particular,
the terms $\frac{\partial \delta p_i}{\partial \overline{\theta}_j}$
and $\frac{\partial \delta q_i}{\partial \overline{\theta}_j}$ in
Eq.~(\ref{dqdp2}) indicate that it is important to account for how
dynamical fluctuations change with $\bar{\theta}_j$.  This is a
crucial piece of information regarding the overall feature of the
dynamical fluctuations.

\subsection{A differential equation describing dynamical fluctuations}

Both Eq~(\ref{dqdp1}) and Eq.~(\ref{dqdp2}) deal with the same time
dependence of $(\mathbf{p},\mathbf{q})$ and hence they should be
consistent with each other. Comparing these two equations term by
term, we arrive at the following equation,
\begin{equation} \label{ZZ1}
\Gamma \left(\begin{array}{c}\delta\mathbf{p}\\
\delta\mathbf{q}\end{array}\right)=\mathbf
{\Sigma}\cdot\frac{d\mathbf{R}}{dt}+\Pi \left(\begin{array}{c}\delta I_1\\
\delta I_2 \\ \vdots \\ \delta I_N
\end{array}\right)+\left(\begin{array}{c}\frac{\partial
\delta \mathbf{p}}{\partial \overline{\theta}_j}\omega_j\\
\frac{\partial \delta \mathbf{q}}{\partial
\overline{\theta}_j}\omega_j
\end{array}\right),
\end{equation}
where
\begin{equation}
\Gamma=\left(\begin{array}{cc}-\frac{\partial^2\overline{H}}{\partial
\overline{\mathbf{q}}\partial\overline{\mathbf{p}}}&-\frac{\partial^2\overline{H}}{\partial
\overline{\mathbf{q}}\partial\overline{\mathbf{q}}}
\\\frac{\partial^2\overline{H}}{\partial \overline{\mathbf{p}}
\partial \overline{\mathbf{p}}}&\frac{\partial^2\overline{H}}{\partial
\overline{\mathbf{p}} \partial \overline{\mathbf{q}}}
\end{array}\right)
\end{equation}
is a $2N\times2N$ matrix;
\begin{equation}
\mathbf{\Sigma}=\left(\begin{array}{c}-\frac{\partial
\overline{\mathbf{p}}}{\partial
\overline{I}_j}\frac{\partial\mathbf{W}}{\partial\overline{\theta}_j}+\frac{\partial
\overline{\mathbf{p}}}{\partial
\overline{\theta}_j}\frac{\partial\mathbf{W}}{\partial
\overline{I}_j}+\frac{\partial \overline{\mathbf{p}}}{\partial
\mathbf{R}}\\
-\frac{\partial \overline{\mathbf{q}}}{\partial
\overline{I}_j}\frac{\partial\mathbf{W}}{\partial\overline{\theta}_j}+\frac{\partial
\overline{\mathbf{q}}}{\partial
\overline{\theta}_j}\frac{\partial\mathbf{W}}{\partial
\overline{I}_j}+\frac{\partial \overline{\mathbf{q}}}{\partial
\mathbf{R}}\end{array}\right)
\end{equation}
is a $2N\times1$ vector along each direction of $\mathbf{R}$; and
\begin{equation}
\Pi=\left(\begin{array}{ccc}\frac{\partial
\overline{\mathbf{p}}}{\partial \overline{\theta}_j}\frac{\partial
\omega_j}{\partial \overline{I}_1}&\frac{\partial
\overline{\mathbf{p}}}{\partial
\overline{\theta}_j}\frac{\partial \omega_j}{\partial \overline{I}_2}&\cdots\\
\frac{\partial \overline{\mathbf{q}}}{\partial
\overline{\theta}_j}\frac{\partial \omega_j}{\partial
\overline{I}_1}&\frac{\partial \overline{\mathbf{q}}}{\partial
\overline{\theta}_j}\frac{\partial \omega_j}{\partial
\overline{I}_2}&\cdots
\end{array}\right)
\end{equation}
is a $2N\times N$ matrix. Substituting Eq.~(\ref{deltadefine}) into
Eq.~(\ref{ZZ1}), one finally obtains an equation for $\delta
\mathbf{p}$ and $\delta \mathbf{q}$ only:
\begin{equation} \label{differential}
\left(\begin{array}{c}\frac{\partial
\delta \mathbf{p}}{\partial \overline{\theta}_j}\omega_j\\
\frac{\partial \delta \mathbf{q}}{\partial
\overline{\theta}_j}\omega_j\end{array}\right)+(\Pi
K-\Gamma)\left(\begin{array}{c}\delta\mathbf{p}\\
\delta\mathbf{q}\end{array}\right)+\mathbf{\Sigma}\cdot\frac{d\mathbf{R}}{dt}=0.
\end{equation}
For a given integrable Hamiltonian, except for those related to
$\delta \mathbf{p}$ and $\delta \mathbf{q}$ and their derivatives,
all the matrices contained in Eq.~(\ref{differential}) are evaluated
at an idealized orbit and hence can be explicitly obtained.

Some remarks are in order. First, Eq.~(\ref{differential}) is not
about evolving the fluctuations
$(\delta\mathbf{q},\delta\mathbf{p})$ at one moment to the next
moment. Instead, it describes, when the system parameters reach the
current configuration $\mathbf{R}$ with a small but nonzero rate
$\frac{d\mathbf{R}}{dt}$, the deviation of the overall shape of one
true orbit from the idealized orbit without dynamical fluctuations,
i.e., the overall deformed orbit in phase space. To our knowledge,
this result is obtained for the first time here. This detailed
description of the dynamical fluctuations can be very useful for
both quantitative and qualitative considerations. The derivation
here is somewhat lengthy because the physical meaning of
$(\overline{\mathbf{I}},\overline{\mathbf{\Theta}})$ in terms of
$({\bf q},{\bf p})$ and hence the idealized orbit itself is changing
as $\mathbf{R}$ varies. Second, consistent with our treatment to the
first order of $\epsilon$, $(\delta\mathbf{p},\delta\mathbf{q})$ is
seen to depend on $\frac{d\mathbf{R}}{dt}$. If
$\frac{d\mathbf{R}}{dt}$ were identically zero, then
$\delta\mathbf{p}=\delta\mathbf{q}=0$ is one possible solution (If
$\delta\mathbf{p}\ne 0$ and $\delta\mathbf{q}\ne 0$ is still the
solution for $\frac{d\mathbf{R}}{dt}=0$, then this solution
describes the relationship between two infinitely close orbits).
 Third, in the absence of the detailed information of
$(\delta \mathbf{p},\delta\mathbf{q})$ for at least one phase space
location, Eq.~(\ref{differential}) alone does not suffice to predict
$(\delta\mathbf{p},\delta\mathbf{q})$ because of its differential
form. As will be discussed later, this implies that in general,
detailed information of the time-dependence of ${\mathbf R}$, e.g.,
its smoothness, can be important for determining the dynamical
fluctuations.
Finally, because the linear Schr\"{o}dinger equation and nonlinear
Gross-Pitaeviskii (GP) equation have an exact canonical structure of
Hamiltonian dynamics \cite{Weinberg,HeslotPRD}, our results here can
be also relevant to quantum adiabatic processes.

If we now consider the mean behavior of
$(\delta\mathbf{p},\delta\mathbf{q})$ along an ideal orbit (denoted
by $\langle\cdot\rangle$), then using the fact that
$(\delta\mathbf{p},\delta\mathbf{q})$ are periodic functions of
$\overline{\mathbf{\Theta}}$, we reduce Eq.~(\ref{differential}) to
\begin{equation} \label{mean}
\left\langle(\Pi
K-\Gamma)\left(\begin{array}{c}\delta\mathbf{p}\\
\delta\mathbf{q}\end{array}\right)\right\rangle+\langle\mathbf{\Sigma}\rangle\cdot\frac{d\mathbf{R}}{dt}=0.
\end{equation}
Because the matrices $\Pi$, $K$, $\Gamma$ vary along the orbit, one
may infer from Eq.~(\ref{mean}) the statistical correlations
$\langle(\Pi K-\Gamma)\delta\mathbf{p}\rangle$ and $\langle(\Pi
K-\Gamma)\delta\mathbf{q}\rangle$, but the mean fluctuations
$\langle \delta\mathbf{p}\rangle$, $\langle
\delta\mathbf{q}\rangle$, or $\langle \delta\mathbf{I}\rangle$
remain unknown.

\section{Emergence of a geometric angle from an adiabatically moving fixed-point solution}
As a direct application of our central result in
Eq.~(\ref{differential}), here we focus on a rather simple case,
where the solution to Hamilton's equation of motion is a fixed point
in phase space if the system parameters are not changing. We denote
the fixed-point solution as $(\overline{\mathbf{p}},
\overline{\mathbf{q}})$, which are of course functions of
$\mathbf{R}$.  Consider now an adiabatic process in which
$\mathbf{R}$ is changing slowly. Then the idealized orbit according
to CAT is just one adiabatically moving fixed point. In addition, at
this fixed point all functions of $(\overline{\mathbf{p}},
\overline{\mathbf{q}})$ are independent of $\overline{\bf \Theta}$
(otherwise they would be time-dependent), thus forcing their
derivatives with respect to $\overline{\bf \Theta}$ to vanish and
making an averaging over $\overline{\bf \Theta}$ [e.g., in
Eq.~(\ref{mean})] unnecessary.
 We therefore obtain
\begin{eqnarray}
{\Pi}&=&0, \\
{\mathbf{\Sigma}}&=&(\frac{\partial \overline{\mathbf{p}} }{\partial
\mathbf{R}}, \frac{\partial \overline{\mathbf{q}}}{\partial
\mathbf{R}})^{T}.
\end{eqnarray}
Using these results we have the following relation from
Eq.~(\ref{differential}):
\begin{equation} \label{mean3}
\left(\begin{array}{c} \delta{\mathbf{p}}\\
\delta {\mathbf{q}}\end{array}\right)= {\Gamma}^{-1}
 \left(\begin{array}{c}\frac{\partial \overline{\mathbf{p}}}{\partial \mathbf{R}}\\
 \frac{\partial \overline{\mathbf{q}}}{\partial \mathbf{R}}\end{array}\right) \cdot \frac{d\mathbf{R}}{dt}.
\end{equation}

Note that the values of $\theta_i$ at a fixed point can be taken as
arbitrary. Hence the fluctuations obtained in Eq.~(\ref{mean3}) do
not have any interesting consequence for the evolution of
$\theta_i$. Furthermore, since the $K$ matrix vanishes at fixed
points (where the action reaches its minimum), one would also arrive
at $\delta{\mathbf{I}}=0$ to the first order of $\epsilon$ even
though $\delta\mathbf{q} \ne 0$ and $\delta\mathbf{p}\ne 0$.

Consider then the coupling of this system with another degree of
freedom, whose canonical coordinates are denoted by $(J,\phi)$. The
total Hamiltonian is assumed to be independent of $\phi$, denoted
$H^{\text{tot}}({\bf p}, {\bf q}, J)$. Because $J$ is a strict
constant of motion and can be regarded as a fixed system parameter
for the motion of $(\mathbf{p},\mathbf{q})$, the expression for
$\delta \mathbf{p}$ and $\delta \mathbf{q}$ in Eq.~(\ref{mean3})
still applies to fixed points in the phase space of
$(\mathbf{p},\mathbf{q})$.  To seek how fluctuations predicted by
Eq.~(\ref{mean3}) may affect the motion in $\phi$,  let us now
examine the angular frequency associated with $\phi$, i.e.,
\begin{eqnarray}
\omega_J({\bf p}, {\bf q}, J)\equiv \frac{\partial
H^{\text{tot}}}{\partial J}.
\end{eqnarray} Clearly, the fluctuations $\delta\mathbf{p}$ and $\delta\mathbf{q}$ will lead to
\begin{eqnarray}
\delta \omega_J(\overline{{\bf p}}, \overline{{\bf q}}, J) =
\frac{\partial \omega_J(\overline{{\bf p}}, \overline{{\bf q}},
J)}{\partial {\bf \overline{p}}}\cdot \delta {\bf p} +
 \frac{\partial \omega_J(\overline{{\bf p}}, \overline{{\bf q}}, J)}{\partial {\bf \overline{q}}}\cdot \delta {\bf q}. \nonumber \\
\end{eqnarray} This fluctuation in  $\omega_J({\bf p}, {\bf q}, J)$ induces an correction to the evolution of $\phi$.
Using Eq.~(\ref{mean3}), one finds an explicit expression for this
correction as follows,
\begin{eqnarray}
 \phi^{\text{corr}} &=&
 \int_0^T \left[ \frac{\partial \omega_{J}}{\partial \overline{\mathbf{p}}}\cdot  \delta\mathbf{p}
  +\frac{\partial \omega_{J}}{\partial \overline{\mathbf{q}}}\cdot  \delta\mathbf{q} \right] \ dt \nonumber \\
 & = &  \oint \left(\begin{array}{cc}\frac{\partial \omega_J}{\partial \overline{\mathbf{p}}}, &\frac{\partial
 \omega_{J}}{\partial \overline{\mathbf{q}}}
 \end{array}\right){\Gamma}^{-1} \left(\begin{array}{c}\frac{\partial
\overline{\mathbf{p}}}{\partial \mathbf{R}}\\ \frac{\partial
\overline{\mathbf{q}}}{\partial \mathbf{R}}\end{array}\right) \cdot
 d\mathbf{R}.
 \label{thetaresult}
\end{eqnarray}
As seen from Eq.~(\ref{thetaresult}), $ \phi^{\text{corr}}$ obtained
above no longer depends on $T$ (so it will not vanish even in the
$\epsilon\rightarrow 0$ or $T\rightarrow +\infty$ limit). Rather, it
depends on the geometry in the parameter space only. $
\phi^{\text{corr}}$ is hence identified as a geometric angle that
arises from the fluctuations in a classical adiabatic process. This
is particularly interesting because here $\delta\mathbf{I}=0$, i.e.,
even when the fluctuations in the original action variables are
vanishing, there can still be a physical effect on another degree of
freedom due to the dynamical fluctuations.

To illustrate the result in Eq.~(\ref{thetaresult}) we have designed
a simple toy model with two degrees of freedom in total.
Specifically, the total Hamiltonian is given by
\begin{equation} \label{Hexample}
H^{\text{tot}}(p_1,q_1; J)=\alpha J+
\frac{1}{2}\left[\left(\frac{p_1^2}{X^2}-J\right)^2+\left(\frac{q_1^2}{Y^2}-J\right)^2\right],
\end{equation}
with $\mathbf{R}=(X>0,Y>0)$, $\phi$ being a cyclic angular
coordinate that forms a canonical pair with $J$, and $\alpha$ being
a free parameter. For the $(p_1,q_1)$ degree of freedom, this system
has a $\mathbf{R}$-dependent fixed point
\begin{eqnarray}
\overline{q}_1&=&\sqrt{\bar{J}}Y; \\ \nonumber
\overline{p}_1&=&\sqrt{\bar{J}}X,
\end{eqnarray} where $\bar{J}$ represents a
conserved value of the variable $J$.

To calculate the fluctuation-induced geometric angle seen in the
evolution of $\phi$, note first
\begin{equation}
\omega_J = \frac{\partial H^{\text{tot}}}{\partial J}=\alpha+
2J-\frac{p_1^2}{X^2}-\frac{q_1^2}{Y^2},
\end{equation} and
\begin{equation}
\frac{\partial\omega_J}{\partial
\overline{p}_1}=-\frac{2\overline{p}_1}{X^2},\
\frac{\partial\omega_J}{\partial
\overline{q}_1}=-\frac{2\overline{q}_1}{Y^2}.
\end{equation}
One may also easily obtain that the matrix ${\Gamma}$ here is just a
$2\times 2$ matrix, i.e.,
\begin{equation}
{\Gamma}_{2\times
2}=\left(\begin{array}{cc}-\frac{\partial^2\overline{H}}{\partial
\overline q_1\partial
\overline{p}_1}&-\frac{\partial^2\overline{H}}{\partial
\overline{q}_1\partial \overline{q}_1}
\\ \frac{\partial^2\overline{H}}{\partial \overline{p}_1 \partial
\overline{p}_1}&\frac{\partial^2\overline{H}}{\partial
\overline{p}_1 \partial \overline{q}_1}
\end{array}\right)=\left(\begin{array}{cc}0&-\frac{4\bar{J}}{Y^2}
\\ \frac{4\bar{J}}{X^2}&0
\end{array}\right);
\end{equation}  and
\begin{equation}
\left(\begin{array}{c}\frac{\partial \overline{p}_1}{\partial \mathbf{R}}\\
\frac{\partial \overline{q}_1}{\partial
\mathbf{R}}\end{array}\right)=\left(\begin{array}{c}\sqrt{\bar{J}}\\
0\end{array}\right)\hat{X}+\left(\begin{array}{c}0\\
\sqrt{\bar{J}}\end{array}\right)\hat{Y},
\end{equation}
where $\hat{X}$ and $\hat{Y}$ are unit vectors along the $X$ and $Y$
coordinates.  Finally, substituting these intermediate results into
Eq.~(\ref{thetaresult}), one finds the fluctuation-induced geometric
angle
\begin{eqnarray}
\phi^{\text{corr}}&=&
 \oint_C \left(\begin{array}{cc}\frac{\partial \omega_J}{\partial \overline{p}_1}&\frac{\partial \omega_J}{\partial
 \overline{q}_1} \end{array}\right)
 {\Gamma}_{2\times 2}^{-1} \left(\begin{array}{c}\frac{\partial \overline{p}_{1}}{\partial \mathbf{R}}\\ \frac{\partial
 \overline{q}_{1}}{\partial \mathbf{R}}\end{array}\right)
 \cdot d\mathbf{R} \nonumber\\
&=&\oint_C \left(\frac{\partial\omega_J}{\partial \overline{p}_1}
\frac{X^2}{4\bar{J}}\sqrt{\bar{J}}\hat{Y},\
-\frac{\partial\omega_J}{\partial
\overline{q}_1}\frac{Y^2}{4\bar{J}}\sqrt{\bar{J}}\hat{X} \right)
\cdot d\mathbf{R} \nonumber \\ &=&\frac{1}{2}\oint_C (Y dX-X dY)
=-\iint\limits_{\partial S=C} dS. \label{phiresult}
\end{eqnarray}
As seen from the above result, here the geometric angle induced by
the fluctuations in the first degree of freedom may be interpreted
as the flux of an effective ``magnetic charge" uniformly distributed
on the $(X,Y)$ plane. The emergence of such a new classical
geometric angle from our simple calculations is hence intriguing.
It should be emphasized that in obtaining $\phi^{\text{corr}}$ in
Eq.~(\ref{phiresult}), we did not seek new action-angle variables
$(\tilde{I}_1,\tilde{\theta}_1)$ and
$(\tilde{I}_2,\tilde{\theta}_2)$ such that $H^{\text{tot}}$ becomes
a function of $\tilde{I}_1$ and $\tilde{I}_2$ only. Indeed it can be
highly complicated in general to find such a new representation due
to the coupling between the two degrees of freedom. This indicates
that $\phi^{\text{corr}}$ here has a different meaning than Hannay's
angle, because it represents a geometrical correction to the $\phi$
evolution, not to the evolution of the yet-to-be-found new angle
variables $\tilde{\theta}_1$ or $\tilde{\theta}_2$.

We also note that our result here is consistent with one of the
found terms in the previous study of the so-called ``nonlinear Berry
phase" based on GP equation~\cite{liu}.  In particular, the GP
equation considered in Ref.~\cite{liu} can be mapped to that of a
classical Hamiltonian with two degrees of freedom, with the
nonlinear eigenstates mapped to classical fixed points (see also
Sec.~IV-B). Adopting our perspective here, the geometric phase
contributed by deviations from nonlinear eigenstates as analyzed in
Ref.~\cite{liu} may be understood as a classical geometric angle due
to intrinsic fluctuations in classical adiabatic processes.  Indeed,
we have checked that if we apply Eq.~(\ref{mean3}) to the model
considered in Ref.~\cite{liu}, then we can obtain a
fluctuation-induced geometric phase term that is identical with a
Berry-phase correction term discovered in Ref.~\cite{liu}. Note
however, the focus of our perspective is on a general description of
the important dynamical fluctuations in a broad class of classical
adiabatic processes. In our fully classical considerations here, a
totally classical geometry angle is shown to arise in a second
degree of freedom that is coupled with the first degree of freedom
(with one adiabatically moving fixed point solution); whereas in
Ref.~\cite{liu}, the emphasis was placed on a quantum adiabatic
evolution context and the main concern is with the sum of one
familiar Berry phase and a fluctuation-induced geometric phase as a
correction.

\section{Discussion}
\subsection{Pollution to Hannay's angle}

As mentioned above, in some early studies about Hannay's angle in
some Hamiltonian systems \cite{Golin1, Golin2, Berry1996Non,adam},
it was numerically found that during an adiabatic process the total
angle change minus the dynamical angle may not be Hannay's angle.
This subtle behavior was connected with dynamical fluctuations in
classical adiabatic processes. Here we exploit our general result of
Eq.~(\ref{differential}) to shed more light on possible pollution to
Hannay's angle.

According to Eq.~(\ref{new-dyna2}) and CAT, the total change in
angle variables in a cyclic adiabatic process is given by
\begin{equation}
\triangle\theta_{i}^{\text{ideal}}(T)=\int_0^{T}\omega_{i}(\overline{\mathbf{I}},\mathbf{R})\
dt - \frac{\partial}{\partial \overline{I}_i} \oint
(\overline{\mathbf{p}}\cdot
\mathbf{\nabla}_{\mathbf{R}}\overline{\mathbf{q}}) \cdot
d\mathbf{R}. \label{standard}
\end{equation}
On the right hand side of  Eq.~(\ref{standard}), the first term is
often called the dynamical angle, and the second term gives Hannay's
angle (upon an average over initial angle variables). We have also
used the notation $\theta_{i}^{\text{ideal}}$ to emphasize that it
is for idealized cases without considering any dynamical
fluctuations.  Indeed, the angular frequency $\omega_{i}$ in
Eq.~(\ref{standard}) is naively assumed to be the one determined by
the idealized and constant action $\overline{\mathbf{I}}$.

However, as suggested by Eq.~(\ref{new-dyna2}), fluctuations in the
action variables $\delta\mathbf{I}$ can then correct the angular
frequency from $\omega_{i}(\overline{\mathbf{I}},\mathbf{R})$ to
$\omega_{i}(\overline{\mathbf{I}},\mathbf{R})+\frac{\partial
\omega_i(\overline{\mathbf{I}},\mathbf{R})}{\partial
\overline{\mathbf{I}}} \cdot\delta\mathbf{I}$. In terms of the
canonical variables $(\mathbf{p},\mathbf{q})$, fluctuations in
$\overline{\mathbf{p}}$ and $\overline{\mathbf{q}}$ will lead to
fluctuations in the angular frequency
\begin{eqnarray}
\delta \omega(\overline{{\bf I}}, {\bf R}) = \frac{\partial
\omega(\overline{{\bf I}}, {\bf R})}{\partial {\bf
\overline{p}}}\cdot \delta {\bf p} +
 \frac{\partial \omega(\overline{{\bf I}}, {\bf R})}{\partial {\bf \overline{q}}}\cdot \delta {\bf q}. \label{deltaw}
\end{eqnarray}
For this reason, the dynamical angle obtained by a time-integral of
the idealized frequency
$\omega_{i}(\overline{\mathbf{I}},\mathbf{R})$, [see
Eq.~(\ref{standard})] should be re-examined with care. In terms of
$\delta\mathbf{p}$ and $\delta\mathbf{q}$, the real change in the
angular variables should be given by
\begin{eqnarray}
\triangle\theta_{i}^{\text{real}}(T)& = &
\triangle\theta_{i}^{\text{ideal}}(T)+\int_0^{T}
\delta\omega_i(\mathbf{I}, \mathbf{R})\ dt \nonumber \\
&=& \triangle\theta_{i}^{\text{ideal}}(T)+\int_0^{T}\frac{\partial
\omega(\overline{{\bf I}}, {\bf R})}{\partial {\bf
\overline{p}}}\cdot \delta
{\bf p}\ dt \nonumber \\
&& +\ \int_0^{T} \frac{\partial \omega(\overline{{\bf I}}, {\bf
R})}{\partial {\bf \overline{q}}}\cdot \delta {\bf q} \ dt .
 \label{standardreal}
\end{eqnarray}
Because $\delta \mathbf{p}$, $\delta \mathbf{q}$ and hence $\delta
\omega$ are of the same order with $\epsilon=|d\mathbf{R}/dt|$, just
like the above fixed-point solution case, the term $\int_0^{T}
\delta\omega_i \ dt$ may not be negligible as it accumulates the
fluctuations $\delta\omega_i(\mathbf{I}, \mathbf{R})$ over an entire
adiabatic process. So the term $\int_0^{T} \delta\omega_i \ dt$
should not be neglected without a clear understanding of the
dynamics. At this point it is also clearer why we only consider
$\delta \mathbf{p}$ and $\delta \mathbf{q}$ to the first order of
$\epsilon$: including higher-order terms are unnecessary because
they will vanish in the $\epsilon\rightarrow 0$ limit.

The correction term $\int_0^{T} \delta\omega_i \ dt$ can hence give
the difference between two objects: the standard Hannay's angle, and
a numerical calculation of a geometric angle based on the expression
of
($\triangle\theta_{i}^{\text{real}}-\int_0^{T}\omega_{i}(\overline{\mathbf{I}},\mathbf{R})\
dt$).  Unfortunately, unless for special fixed-point solution cases
analyzed above, we in general cannot determine the fluctuations
$\delta \mathbf{p}$ and $\delta\mathbf{q}$ from the differential
equation in Eq.~(\ref{differential}). In particular, $\delta
\mathbf{p}$ and $\delta\mathbf{q}$ can only be determined if we have
information about them for at least one given $\mathbf{\Theta}$ (as
the input).  Therefore, without some detailed information of an
adiabatic process, e.g., the detailed dependence of adiabatic
parameter $\mathbf{R}(t)$ on time, information about $\delta
\mathbf{p}$ and $\delta \mathbf{q}$ is not available in general.

To see more clearly, let us discretize the adiabatic process by
dividing one adiabatic process into many time intervals
$t_1,t_2,\cdots$, during each of which $\mathbf{R}=\mathbf{R}_j$,
followed by a jump onto the next value $\mathbf{R}_{j+1}$ after the
temporal interval $t_j$ (different time intervals and different
choices for $\mathbf{R}_j$ define different adiabatic processes with
different details). Note that even for a continuous adiabatic
process, this discretized version is rather typical in numerical
simulations (as the discretized time steps decrease, the simulated
dynamics approaches a continuous process). Now for each point
$\mathbf{R}_j$,  we may use Eq.~(\ref{differential}) to describe the
dynamical fluctuations, but Eq.~(\ref{differential}) is dependent on
$\mathbf{R}_j$. For a particular segment where
$\mathbf{R}=\mathbf{R}_j$,  the angle variable $\mathbf{\Theta}$
changes rapidly. Obviously, different timing for the next jump will
result in different initial values of $\mathbf{\Theta}$ for next
segment $\mathbf{R}=\mathbf{R}_{j+1}$, leading to another initial
condition for the differential equation (\ref{differential})
associated with ${\bf R}=\mathbf{R}_{j+1}$. This process then
continues.
According to Eq.~(\ref{deltaw}), $\delta\omega$ and thus the
correction term $\int_0^{T} \delta\omega_i \ dt$ will then depend on
great details of a particular adiabatic process.  It is for this
reason that the correction term $\int_0^{T} \delta\omega_i \ dt$ is
identified as  ``pollution" to Hannay's angle, with the latter
independent of how an adiabatic process is implemented.   Analysis
here also makes it clearer that the fixed-point solution case in
Sec.~III is special because a definite prediction about fluctuations
can be made therein.

It is also worth noting that, according to Eq.~(\ref{deltaw}), the
pollution vanishes if the angular frequency $\omega_i$ does not
depend on the action $\mathbf{I}$. This is the case in a linear
system such as a harmonic oscillator.

\subsection{``Pollution" to adiabatic phase evolution in a two-mode BEC model}

Finally, we propose to use a two-mode GP equation to study pollution
to a geometric phase associated with quantum adiabatic cycles, thus
making a connection between our theoretical considerations here and
a reachable experimental context. In particular, there are a number
of possibilities to experimentally realize a two-mode BEC. For
example, one may consider a BEC in a double-well potential, or a BEC
in an optical lattice occupying two bands \cite{BECexpe}. On the
mean-field level, a two-mode BEC can be described by the following
GP equation ($\hbar=1$)
\begin{eqnarray} \label{GP} \nonumber
&i\frac{d}{dt}\left(\begin{array}{c}a\\b\end{array}\right)={H}_{\text{GP}}\left(\begin{array}{c}a\\b\end{array}\right)\\
&=\frac{1}{2}\left(\begin{array}{cc}\gamma+c(|b|^2-|a|^2)&\Delta\\
\Delta&-\gamma-c(|b|^2-|a|^2)\end{array}\right)\left(\begin{array}{c}a\\b\end{array}\right), \nonumber  \\
\end{eqnarray}
where $\gamma$ denotes an energy bias between the two modes, $|a|^2$
and $|b|^2$ (with $|a|^2+|b|^2=1$) represent occupation
probabilities of the two modes, $c$ gives the self-interaction
strength, and $\Delta$ denotes the coupling between the two modes.
We can consider, for example, the two parameters $\gamma$ and
$\Delta$ to implement an adiabatic cyclic process.

The dynamics described by the above GP equation can be translated
into Hamiltonian dynamics. In particular,  let $p=\phi_a-\phi_b$,
$q=|a|^2$, $a=|a|e^{i\phi_a}$, $b =|b|e^{i\phi_b}$, then apart from
an overall phase parameter $\phi_b$, Eq.~(\ref{GP}) leads to
\begin{eqnarray} \label{classical} \nonumber
&\frac{dp}{dt}=-\frac{\partial H}{\partial q}, \nonumber \\
&\frac{dq}{dt}=\frac{\partial H}{\partial p}.
\end{eqnarray}
where
\begin{eqnarray} \label{CHamiltonian} \nonumber
H&=&\Delta\sqrt{q(1-q)}+\frac{\gamma}{2}(2q-1)-\frac{c}{4}(2q-1)^2.
\end{eqnarray}
It is also straightforward to find that the evolution of $\phi_b$
obeys
\begin{eqnarray}
\frac{d\phi_b}{dt}=i(\sqrt{q}e^{-ip},\sqrt{1-q})\frac{d}{dt}\left(\begin{array}{c}\sqrt{q}e^{ip}\\
\sqrt{1-q}\end{array}\right)-H-\Lambda, \label{phaseevo}
\end{eqnarray}
where
\begin{eqnarray}
 \Lambda&=&-\frac{c}{4}(2q-1)^2.
\end{eqnarray}
It is seen that the evolution of the overall phase $\phi_b$ is
determined by, but will not have a back action on, the classical
trajectories determined by $H$ in Eq.~(\ref{CHamiltonian}). In this
sense, the $\phi_b$ parameter plays a similar role as the $\phi$
parameter in Sec.~III.

It is now clear that our general result of dynamical fluctuations in
classical adiabatic processes can be directly relevant to
understanding the adiabatic evolution of a two-mode BEC system. If
the adiabatic process starts from a stationary state of the GP
equation, then the dynamics is just about
 an adiabatically evolving fixed-point solution of the Hamiltonian in
Eq.~(\ref{CHamiltonian}).  As shown earlier (see also
Ref.~\cite{liu}), in this case a definite prediction can be made
about how accumulation of dynamical fluctuations can eventually lead
to a geometry-like correction to $\phi_b$. Consider now a
superposition state of two stationary states of the above two-mode
GP equation as the initial state of an adiabatic process. This case
then corresponds to a classical adiabatic process with
non-fixed-point solutions.  As indicated by Eq.~(\ref{phaseevo}),
dynamical fluctuations can now affect the evolution of the
adiabatically evolving phase $\phi_b$, in an unpredictable way if we
do not know the details of the adiabatic process. Pollution to the
quantum phase $\phi_b$ hence emerges. Interestingly, in the same
context,  how $\phi_b$ may develop an adiabatic geometric phase for
general superpositions of stationary states
 was already considered in Ref.~\cite{WuPRL2005} without considering dynamical fluctuations.
It is hence of interest to numerically or even experimentally
examine the actual pollution due to the accumulation of dynamical
fluctuation effects
 in such type of quantum adiabatic processes.

\section{Summary}
To summarize, we have obtained a general description of the
intrinsic dynamical fluctuations in classical adiabatic processes
associated with integrable systems. These fluctuations are typically
neglected by the conventional classical adiabatic theorem. The
dynamical fluctuations are described in this work in terms of
deviations from idealized adiabatic trajectories.  As an
application, we have shown how a new kind of classical geometric
phase may emerge using an explicit example with an adiabatically
evolving fixed-point solution.  We then discussed the origin of the
pollution to Hannay's angle and proposed to use a two-mode BEC
system to further study possible fluctuation-induced pollution to
one type of quantum adiabatic evolution described on a mean-field
level.

{\bf Acknowledgement} The work of Q.Z. and C.H was supported by
National Research Foundation and Ministry of Education, Singapore
(Grant No. WBS: R-710-000-008-271) and by the National Natural
Science Foundation of China (Grant No. 11105123).

\vspace{0.5cm}

\appendix
\section{On the derivation of Eq.~(\ref{dqdp2})}
Here we present some necessary details in deriving
Eq.~(\ref{dqdp2}). The summation convention by using repeated
indices is also adopted here. First, by definition we have
$p_i=\overline{p}_i+\delta p_i$, and hence
\begin{eqnarray}
\frac{\partial p_i}{\partial \mathbf{R}}=\frac{\partial
\overline{p}_i}{\partial \mathbf{R}}+\frac{\partial
\delta{p}_i}{\partial \mathbf{R}}. \end{eqnarray} As a second step,
let us expand the expressions $\frac{\partial p_i}{\partial I_j}$
and $\frac{\partial p_i}{\partial \theta_j}$ around $\frac{\partial
\overline{p}_i}{\partial \overline{I}_j}$ and $\frac{\partial
\overline{p}_i}{\partial \overline{\theta}_j}$, to the first order
of $\delta {\bf I}$ and $\delta{\bf \Theta}$, leading to
\begin{eqnarray}
\frac{\partial p_i}{\partial I_j}=\frac{\partial
\overline{p}_i}{\partial \overline{I}_j}+\frac{\partial^2
\overline{p}_i}{\partial \overline{I}_j \partial
\overline{I}_k}\delta I_k+\frac{\partial^2 \overline{p}_i}{\partial
\overline{I}_j
\partial \overline{\theta}_k}\delta \theta_k
\end{eqnarray}  and
\begin{eqnarray}
\frac{\partial p_i}{\partial \theta_j}=\frac{\partial
\overline{p}_i}{\partial \overline{\theta}_j}+\frac{\partial^2
\overline{p}_i}{\partial \overline{\theta}_j \partial
\overline{I}_k}\delta I_k+\frac{\partial^2 \overline{p}_i}{\partial
\overline{\theta}_j \partial \overline{\theta}_k}\delta \theta_k.
\end{eqnarray} To proceed further we shall use the obvious two relations
\begin{eqnarray}
\frac{d p_i}{dt}&=&\frac{\partial p_i}{\partial
\mathbf{R}}\frac{d\mathbf{R}}{dt}+ \frac{\partial p_i}{\partial
I_j}\frac{d I_j}{dt}+\frac{\partial p_i}{\partial
\theta_j}\frac{d\theta_j}{dt}; \\
\omega_j(I,\mathbf{R})&=&\omega_j(\bar{\mathbf{I}},
\mathbf{R})+\frac{\partial \omega_j}{\partial \overline{I}_k}\delta
I_k.
 \end{eqnarray} Substituting Eqs.~(1),
(2), (A1), (A2), and (A3) into Eq.~(A4), we find
\begin{eqnarray}
\nonumber \frac{d p_i}{dt}&=&\frac{\partial \overline{p}_i}{\partial
\mathbf{R}}\frac{d\mathbf{R}}{dt}-\frac{\partial
\overline{p}_i}{\partial \overline{I}_j}\frac{\partial \mathbf{W}
}{\partial \overline{\theta}_j}\cdot\frac{d\mathbf{R}}{dt} \nonumber \\
&& + \frac{\partial \overline{p}_i}{\partial
\overline{\theta}_j}\left[\frac{\partial \mathbf{W}}{\partial
\overline{I}_j}\cdot
\frac{d\mathbf{R}}{dt}+\omega_j(\bar{\mathbf{I}},
\mathbf{R})+\frac{\partial \omega_j}{\partial \overline{I}_k}\delta
I_k\right] \nonumber \\
&& +\left(\frac{\partial^2 \overline{p}_i}{\partial
\overline{\theta}_j
\partial \overline{I}_k}\delta I_k+\frac{\partial^2
\overline{p}_i}{\partial \overline{\theta}_j
\partial \overline{\theta}_k}\delta \theta_k \right)\omega_j(\bar{\mathbf{I}},
\mathbf{R}), \\ \end{eqnarray} with all the terms of the second
order of $\frac{d\mathbf{R}}{dt}$ or higher neglected. Note that
consistent with our final result, we have assumed that $(\delta {\bf
p}, \delta{\bf q})$ or $(\delta {\bf I},\delta{\bf\Theta})$ are of
the first order of $\epsilon\equiv \frac{d \mathbf{R}}{dt}$.

As the last step we use the relation
\begin{eqnarray}
\frac{\partial^2 \overline{p}_i}{\partial \overline{\theta}_j
\partial \overline{I}_k}\delta I_k+\frac{\partial^2
\overline{p}_i}{\partial \overline{\theta}_j
\partial \overline{\theta}_k}\delta \theta_k &=&
\frac{\partial}{\partial
\overline{\theta}_j}\left(\frac{\partial\overline{p}_i}{\partial
\overline{I}_k}\delta I_k+\frac{\partial\overline{p}_i}{\partial
\overline{\theta}_k}\delta \theta_k\right) \nonumber \\
&=&\frac{\partial \delta p_i}{\partial
\overline{\theta}_j}.\end{eqnarray} Plugging this simple relation
into Eq.~(A6), we obtain the first equality in Eq.~(\ref{dqdp2}).
The second equality in Eq.~(\ref{dqdp2}) can be obtained in the same
manner.
\newpage






\end{document}